\documentclass[fleqn]{2017SCGE}
\usepackage{graphicx}
\usepackage{dcolumn}
\usepackage{bm}
\usepackage{geometry}
\usepackage{hyperref}
\usepackage{amsmath}
\usepackage{color}

\begin{document}

\ensubject{subject}

\ArticleType{Article}
\Year{2018}
\Month{December}
\Vol{62}
\No{1}
\DOI{10.1007/s11432-016-0037-0}
\ArtNo{000000}
\ReceiveDate{November 20, 2018}
\AcceptDate{December 6, 2018}

\title{Condensation of Eigen Microstate in Statistical Ensemble and Phase Transition}{Condensation of Eigen Microstate in Statistical Ensemble and Phase Transition}

\author[1,3]{Gaoke Hu}{}%
\author[3]{Teng Liu}{}
\author[5]{Maoxin Liu}{}
\author[4]{Wei Chen}{}%
\author[1,2,3]{Xiaosong Chen}{{chenxs@bnu.edu.cn }}

\AuthorMark{Hu G K}

\AuthorCitation{Hu G K, Liu T, Liu M X, et al}


\address[1]{Institute of Theoretical Physics, Key Laboratory of Theoretical Physics, Chinese Academy of Sciences, P.O. Box 2735, Beijing, 100190, China}
\address[2]{School of Systems Science, Beijing Normal University, Beijing 100875, China}
\address[3]{School of Physical Sciences, University of Chinese Academy of Science,  No. 19A Yuquan Road, Beijing 100049, China}
\address[4]{State Key Laboratory of Multiphase Complex Systems, Institute of Process Engineering, Chinese Academy of Sciences, Beijing 100190, China}
\address[5]{State Key Laboratory of Information Photonics and Optical Communications $\&$ School of Science, Beijing University of Posts and Telecommunications, \\Beijing 100876, China}


\abstract{In a statistical ensemble with $M$ microstates, we introduce an $M \times M$ correlation matrix with the correlations between microstates as its elements. Using eigenvectors of the correlation matrix, we can define eigen microstates of the ensemble. The normalized eigenvalue by $M$ represents the weight factor in the ensemble of the corresponding eigen microstate. In the limit $M \to \infty$, weight factors go to zero in the ensemble without localization of microstate. The finite limit of weight factor when $M \to \infty$ indicates a condensation of the corresponding eigen microstate. This indicates a phase transition with new phase characterized by the condensed eigen microstate. We propose a finite-size scaling relation of weight factors near critical point, which can be used to identify the phase transition and its universality class of general complex systems. The condensation of eigen microstate and the finite-size scaling relation of weight factors have been confirmed by the Monte Carlo data of one-dimensional and two-dimensional Ising models.}



\maketitle


\begin{multicols}{2}
\section{Introduction}\label{section1}
In statistical physics, the concept of ensemble in phase space serves as a starting point. An ensemble in phase space is composed of the microstates of system under some thermodynamic conditions. Thermodynamic quantities of system can be obtained by the ensemble-average with the summation over all microstates of the ensemble.  

Three important thermodynamic ensembles were defined by J. W. Gibbs \cite{Gibbs}. They are the micro-canonical ensemble, the canonical ensemble, and the grand canonical ensemble. The micro-canonical ensemble is under the\Authorfootnote 

\noindent thermodynamic conditions that the number of particles $N$, the volume $V$, and the total energy $E$ of system are fixed. In the canonical ensemble, 
the energy of system is not known exactly. In place of the energy, the temperature $T$ is specified. In the grand canonical ensemble, neither the energy nor particle number are fixed. In their place, the temperature $T$ and chemical potential $\mu$ are specified.

In computer simulations or experimental investigations of complex systems under some conditions, snapshots of system can be taken. From these snapshots, we can obtain at first microstates and then a statistical ensemble of the system. In this paper, we study the correlations between microstates in the statistical ensemble. With the correlations between microstates as its elements, we can get a correlation matrix of microstate in the statistical ensemble. Using the eigenvectors of the correlation matrix, the eigen microstates of the statistical ensemble can be defined. The normalized eigenvalues by the number of microstate represent the weight factors of the corresponding eigen microstates in the ensemble. The distribution of eigen microstate in the statistical ensemble can be described by all weight factors.

Our paper is organized as follows. In Section 2, we introduce the correlation between microstate and a correlation matrix. Using its eigenvectors, the eigen microstates of the ensemble are calculated. In one-dimensional and two-dimensional Ising models, their eigen microstates and eigenvalues are studied using the Monte Carlo (MC) simulations. In Section 3, we propose a finite-size scaling relation of the weight factors near critical point, which is confirmed by MC simulation data of Ising models. We make some conclusions finally in Section 4.

\section{Eigen microstates of statistical ensemble and condensation}\label{sec:2}
We consider an Ising model with the Hamiltonian
\begin{eqnarray}
\label{hamiltion}
H=-\sum_{\langle i,j \rangle} J_{ij} S_i S_j  \;,
\end{eqnarray}
where $S_i = \pm 1$ is the spin on site $i$ and $J_{ij}$ is the interaction between spins $i$ and $j$. For the Ising model with $N$ spins, a microstate $I$ of system can be described by a vector with $N$ components as
\begin{eqnarray}
{\pmb A}^I = \frac{1}{\sqrt{N}}\left[ \begin{array}{c} S^I_{1} \\ S^I_{2} \\ \vdots \\S^I_{N}\end{array} \right]\;,
\end{eqnarray}
which is normalized and $|{\pmb A}^I|^2=\left[{\pmb A}^I\right]^T \cdot {\pmb A}^I=1$. The total energy of system at microstate ${\pmb A}^I$ can be written as $H^I= - N \left[{\pmb A}^I\right]^T \cdot \hat{\pmb J} \cdot {\pmb A}^I$ with the interaction matrix $\hat{\pmb J}$ defined by  $J_{ij}$.

At temperature $T$, the microstate ${\pmb A}^I$ has a probability 
\begin{eqnarray}
p({\pmb A}^I)=\frac{1}{Z} e^{- H^I/k_B T}\;,
\end{eqnarray}
where $Z=\sum_{I} e^{-H^I/k_B T}$ and $k_B$ is the Boltzmann constant. In MC simulations,  different microstates of system can be  sampled with the probability factor. $M$ microstates from simulation are taken to build up an ensemble. The correlation between microstates $I$ and $J$ is defined as
\begin{eqnarray}
\label{cmatrix}
C_{IJ}=\left[{\pmb A}^I\right]^T \cdot {\pmb A}^J\;.
\end{eqnarray}
If ${\pmb A}^J={\pmb A}^I$,  $C_{IJ}=1$. When ${\pmb A}^J=-{\pmb A}^I$, $C_{IJ}=-1$. We have $-1 \le C_{IJ} \le 1$ in general. 

With $C_{IJ}$ as its elements, an $M\times M$ correlation matrix ${\pmb C}$ is obtained. We suppose that ${\pmb C}$ has $M$ eigenvectors ${\pmb b}_1, {\pmb b}_2,\cdots,{\pmb b}_M$ with associated eigenvalues $\lambda_1, \lambda_2,\cdots,\lambda_M$. We arrange all eigenvalues in the order  $\lambda_1 \ge \lambda_2 \cdots \ge \lambda_M$. There is a relation
\begin{eqnarray}
\label{eigen}
{\pmb C} {\pmb b}_I = \lambda_I {\pmb b}_I, \; I=1,2,...,M,
\end{eqnarray}
where
\begin{eqnarray}
{\pmb b}_I = \left[ \begin{array}{c} b_{1I} \\ b_{2I} \\ \vdots \\b_{MI}\end{array} \right]\;.
\end{eqnarray}
The normalized eigenvectors are orthogonal each other and follow the relation
\begin{eqnarray}
\label{ortho}
{\pmb b}_I^T \cdot {\pmb b}_J =\sum_{l=1}^M b_{lI} b_{lJ}=\delta_{I,J}\;,
\end{eqnarray}
where $\delta_{I,J}$ is the Kronecker delta. The trace of the correlation matrix is ${\rm tr}\left[{\pmb C}\right]=\sum_{I=1}^M \lambda_I = M$.

From the $M$ eigenvectors, we can define an $M\times M$ matrix
\begin{eqnarray}
{\pmb U}=\left[ {\pmb b}_1 {\pmb b}_2\cdots{\pmb b}_M \right]
\end{eqnarray}
with elements $U_{IJ}= b_{IJ}$. ${\pmb U}$ is an orthogonal matrix and satisfies the condition ${\pmb U}^T \cdot {\pmb U} = {\pmb U} \cdot {\pmb U}^T = {\pmb I}$. After the ${\pmb U}$ transformation of the correlation matrix, we have  ${\pmb U}^T \cdot {\pmb C} \cdot {\pmb U} = {\pmb \Lambda}$, where ${\pmb \Lambda}$ is a diagonal matrix with elements $\Lambda_{IJ} = \lambda_I \delta_{I,J}$.

Using the components of an eigenvector ${\pmb b}_I$, we can introduce an eigen microstate
\begin{eqnarray}
\label{mode}
{\pmb E}^I = \sum_{L=1}^{M} b_{LI} {\pmb A}^L,\;I=1,2,\cdots,M\;,
\end{eqnarray}
which satisfies the relation $|{\pmb E}^I|^2=\left[{\pmb E}^I\right]^T \cdot {\pmb E}^I=\lambda_I$. The correlation between eigen microstates $I$ and $J$ is 
\begin{eqnarray}
\label{ortho_lambda}
C_{IJ}^E=
\left[{\pmb E}^I\right]^T \cdot {\pmb E}_J =\sum_{l,m}^M b_{lI} C_{lm} b_{mJ}=\lambda_I \delta_{I,J}\;.
\end{eqnarray}
Therefore, the correlation matrix ${\pmb C}^E$ is diagonal and there is no correlation between eigen microstates. 

In the ensemble consisting of original microstates, all microstates have the same weight. The weight factor of microstate $I$ 
\begin{eqnarray}
   w_I =C_{II}/M=1/M\;, 
\end{eqnarray}
which satisfies the normalization condition $\sum_{I=1}^M w_I =1$.

In the ensemble consisting of eigen microstates, different microstates have different weights. We define the weight factor of eigen microstate $I$ as
\begin{eqnarray}
   w_I^E =C_{II}^E/M=\lambda_I/M\;,
\end{eqnarray}
which satisfies a normalization condition $\sum_{I=1}^M w_I^E =1$.  

In an ensemble without localization of microstate, all weight factors $w_I^E \to 0 $ in the limit $M \to \infty$.
If the largest weight factor $w_1^E$ becomes finite in the limit $M \to \infty$, this indicates a condensation of eigen microstate ${\pmb E}^1$ in the ensemble. This condensation of microstate is similar to the Bose-Einstein condensation \cite{Pitaevski2003Bose}. Now system has a phase transition with the new phase characterized by the eigen microstate ${\pmb E}^1$. 

The eigenvalue $\lambda_I$ depends on $T$, $N$, and $M$ and $\lambda_I = \lambda_I (T,N,M)$. In the limit $M \to \infty$ at fixed $x=I/M$, an eigenvalue function $\lambda (T,N,x) \equiv \lim_{M \to \infty} \lambda_I (T,N,M)$ is obtained. The normalized condition of eigenvalue function is
\begin{eqnarray}
    \int_0^1 \lambda (T,N, x) dx =1\;.
\end{eqnarray}
In the limit $M \to \infty$, the finite $w_1^E$ implies that the eigenvalue function $\lambda (T,N, x) \to \infty $ when $x \to 0$.

Original microstates can be expressed by eigen microstates as
\begin{eqnarray}
\label{eigenmode}
{\pmb A}^I = \sum_{J=1}^{M} b_{JI} {\pmb E}^J, \;I=1,2,...,M\;.
\end{eqnarray}
Since $|{\pmb E}^I|^2=\lambda_I$, a normalized eigen microstate $\bar{\pmb E}^I = \lambda_I^{-1/2} {\pmb E}^I$ is introduced. With normalized eigen microstates, We have 
\begin{eqnarray}
\label{eigenmode1}
{\pmb A}^I = \sum_{J=1}^{M} b_{JI} \lambda_J^{1/2} \bar{\pmb E}^J, \;I=1,2,...,M\;.
\end{eqnarray}

In the ensemble of original microstate, the magnetization of system can be calculated as
\begin{eqnarray}
\left < m \right > = \frac{1}{M}\sum_{I=1}^{M} m_I \;,
\end{eqnarray}
where 
\begin{eqnarray}
  m_I = \frac{1}{\sqrt{N}}\sum_{i=1}^N A_i^I  
\end{eqnarray}
is the magnetization of original microstate $I$.

Equivalently, we can write the magnetization of system according to Eq. \eqref{eigenmode} as
\begin{eqnarray}
\label{relation}
\left < m \right > = \sum_{J=1}^M \bar{b}_{J} \left[w_J^E\right]^{1/2}  m_J^e\;,
\end{eqnarray}
where
\begin{eqnarray}
    m_J^e &=& \frac{1}{\sqrt{N}}\sum_{i=1}^N \bar{E}_i^J\;,\\
   \bar{b}_J &=& \frac{1}{\sqrt{M}} \sum_{I=1}^M b_{JI}\;.
\end{eqnarray}
Other thermodynamics quantities of system can be calculated in a similar way.

In the following, we will study the eigen microstates and their weight factors in the statistical ensembles of one-domensional and two-dimensional Ising models.

\subsection{One-dimensional Ising model}\label{sec:3}
The one-dimensional (1d) Ising model with the nearest-neighbour interaction has the Hamiltonian
\begin{eqnarray}
    H= - J \sum_{i=1}^{N} S_i S_{i+1} \;.
\end{eqnarray}
Under the periodic boundary condition $S_{N+1}=S_1$, this model can be solved exactly \cite{Baxter}. In the thermodynamic limit $N \to \infty$, its correlation function has an exponential form
\begin{eqnarray}
\left< S_i S_j \right> = \exp \left( - |x_i -x_j|/\xi \right) 
\end{eqnarray}
with the correlation length
\begin{eqnarray}
\label{xi1}
    \xi=\tilde{a} \left(\ln \left[ \coth(1/T^*) \right]\right)^{-1}\;,
\end{eqnarray}
where $\tilde{a}$ is the lattice spacing and $T^*=k_B T/J$. In a finite 1d-Ising chain with periodic boundary condition, the system has susceptibility \cite{Chen2000}
\begin{eqnarray}
\label{xi1d}
    \chi (T^*, N) = \left( \frac{1+e^{-\tilde{a}/\xi}}{1-e^{-\tilde{a}/\xi}} \right) \left( \frac{1-e^{-L/\xi}}{1+e^{-L/\xi}}\right)\;,
\end{eqnarray}
where $L=N \tilde{a}$. The correlation length $\xi \to \infty$ when $T \to 0 $. Although it has been well acknowledged that there is no phase transition in such an 1d-Ising model, we can still consider the zero temperature as a critical point since the correlation length diverges. The thermodynamic quantities of 1d-Ising model should have similar critical behaviors above $T_c$ as that of $d$-dimensional Ising model with $d \ge 2$. Near $T=0$, the susceptibility satisfies asymptotically a finite-size scaling relation
\begin{eqnarray}
    \chi (T^*,N) = \left( L/\tilde{a} \right)^{\gamma/\nu} f_\chi (L/\xi)
\end{eqnarray}
with $\gamma/\nu =1$\cite{Baxter} and the finite-size scaling function 
\begin{eqnarray}
f_\chi (x)= \frac{2}{x} \cdot \frac{1-e^{-x}}{1+e^{-x}}\;.    
\end{eqnarray}
Using the hyperscaling relation $d-2\beta/\nu = \gamma/\nu$ for 1d-Ising model, we obtain $\beta=0$, which is in agreement with that of Ref. \cite{Baxter}.

\begin{figure}[H]
\centering
\includegraphics[scale=0.8]{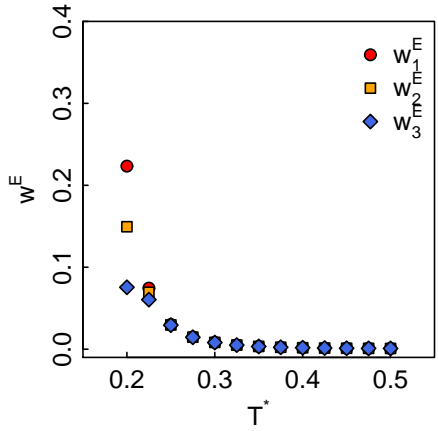}
\caption{The largest three normalized eigenvalues of 1d-Ising model with $N=10^5$ spins.}
\label{fig:eigenvalues1}
\end{figure}

We simulate the microstates of 1d-Ising model using the Wolff algorithm \cite{Wolff1989}. Simulations are started with all spins aligned. To get the microstates in equilibrium, the first $10^4$ microstates are not used. The subsequent microstates are chosen at an interval of 205 MC steps to keep their independence. Using the microstates obtained, we can get the correlation matrix ${\pmb C}$. With the eigenvectors and eigenvalues of ${\pmb C}$, we can obtain eigen microstates according to Eq.  \eqref{mode}.

The $M$-dependence of eigenvalue has been studied.  At $M \approx 10^{4}$, the $M$-dependence of the largest three normalized eigenvalues can be neglected. We show the largest three normalized eigenvalues at $M = 2 \times 10^4$ in Fig. \ref{fig:eigenvalues1}.

We can see from Fig. \ref{fig:eigenvalues1} that the weight factor become finite with the decrease of temperature. This indicates that there will be a phase transition. To identify the new phase of system, the corresponding eigen microstates should be studied. 

\begin{figure}[H]
\centering
\includegraphics[scale=0.5]{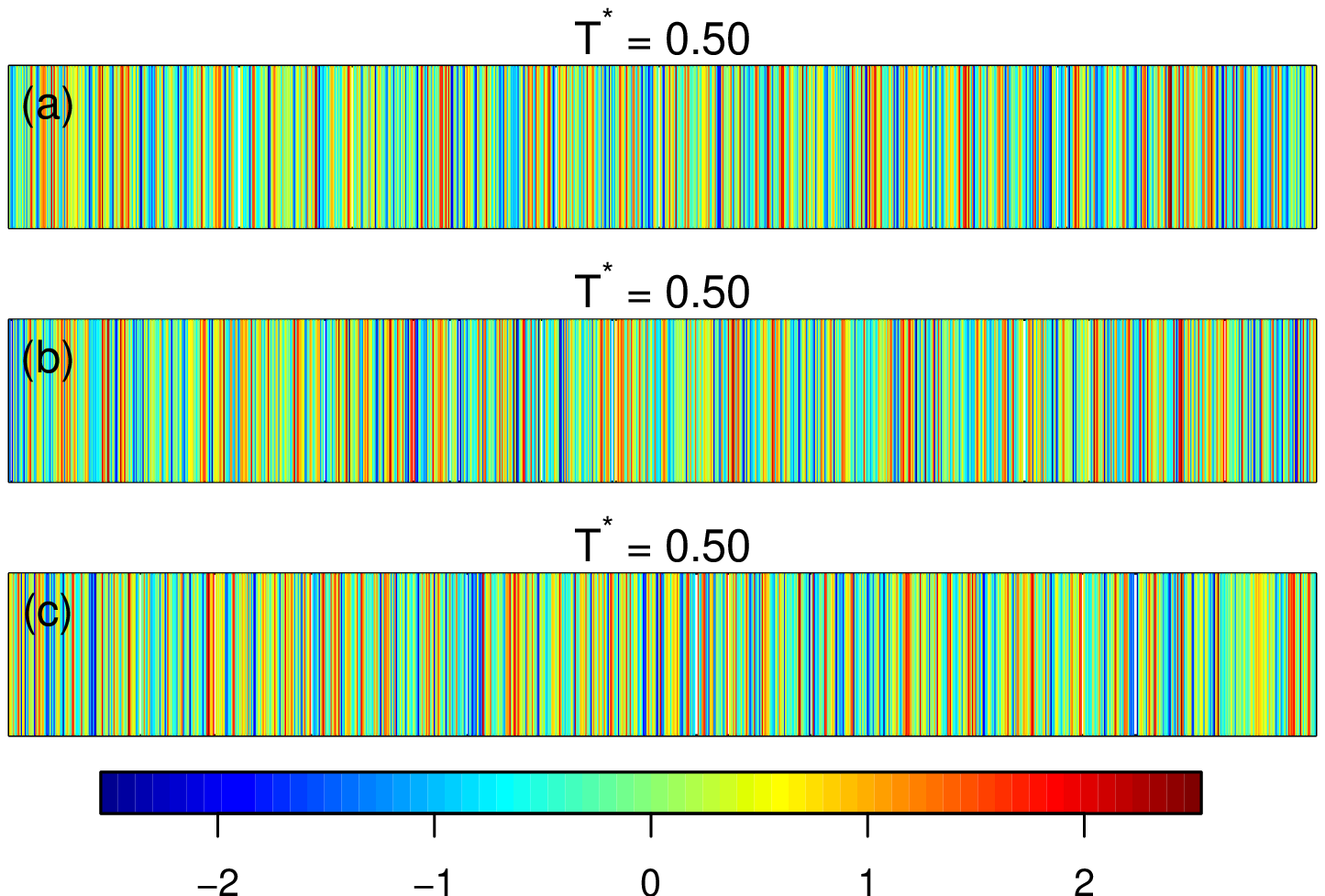}
\caption{The largest  eigen microstate $\sqrt{N}\bar{\pmb E}^1$(a), the second largest eigen microstate $\sqrt{N}\bar{\pmb E}^2$ (b), and the third largest eigen microstate $\sqrt{N}\bar{\pmb E}^3$ (c) of 1d-Ising model.}
\label{fig:eigen microstates5}
\end{figure}

At $T^*=0.5$, the eigen microstates of the largest three eigevalues are shown in Fig. \ref{fig:eigen microstates5}. In these eigen microstates, spin clusters are of micro scales and distributed with alternate orientation in the real space.

\begin{figure}[H]
\centering
\includegraphics[scale=0.5]{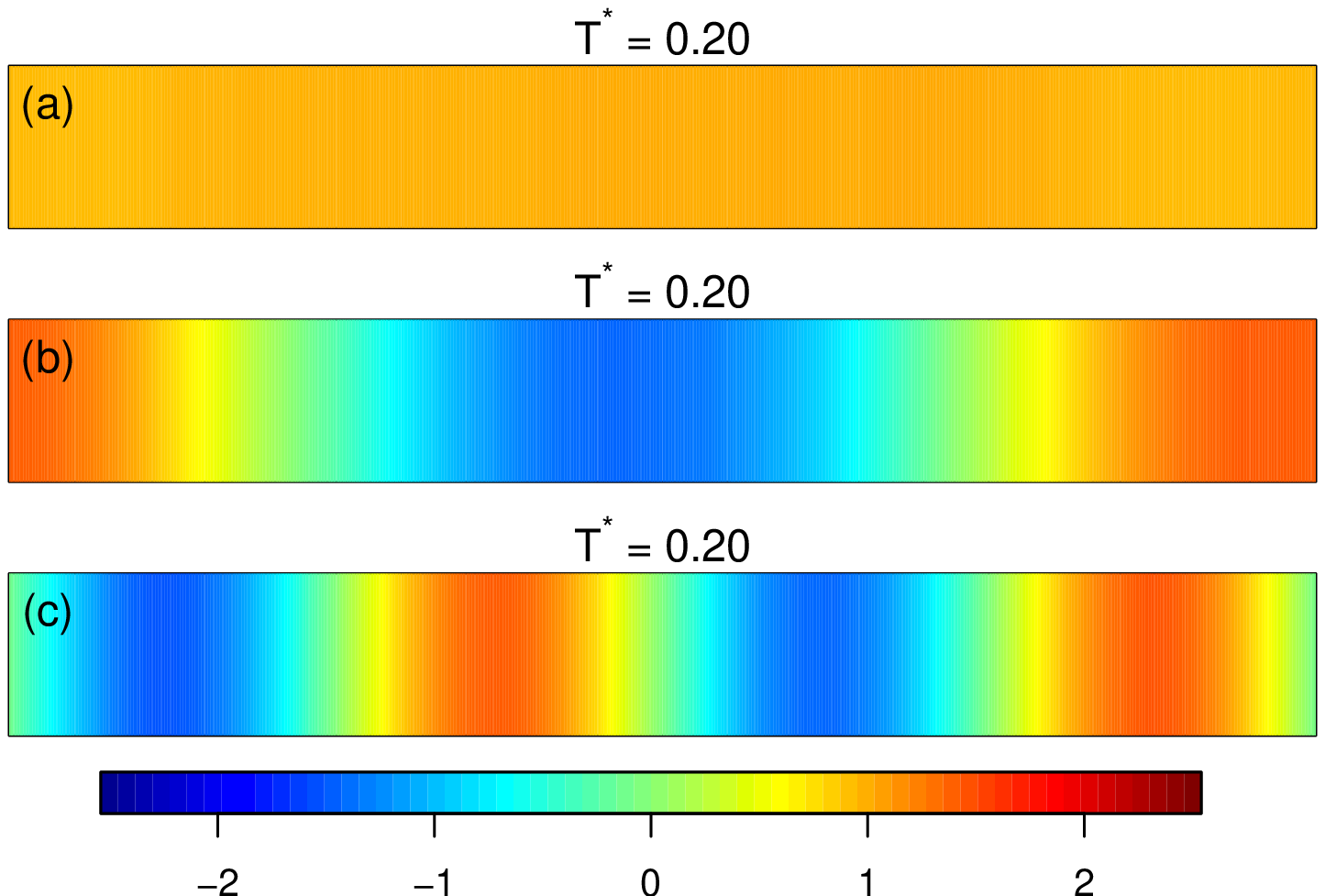}
\caption{The largest  eigen microstate $\sqrt{N}\bar{\pmb E}^1$(a), the second largest eigen microstate $\sqrt{N}\bar{\pmb E}^2$ (b), and the third largest eigen microstate $\sqrt{N}\bar{\pmb E}^3$ (c) of 1d-Ising model.}
\label{fig:eigen microstates2}
\end{figure}

The largest three eigen microstates at $T^*=0.2$ are presented in Fig. \ref{fig:eigen microstates2}. The sizes of clusters are comparable to that of system. Only one cluster exists in the largest eigen microstate. The second largest eigen microstate has two clusters with opposite orientation. In the third largest eigen microstate, there are four clusters with alternate orientations.

For an overview of the weight distribution of eigen microstate, we define the cumulant
\begin{eqnarray}
    c(m) = \sum_{I=1}^m w_I^E\;.
\end{eqnarray}

\begin{figure}[H]
\centering
\includegraphics[scale=0.6]{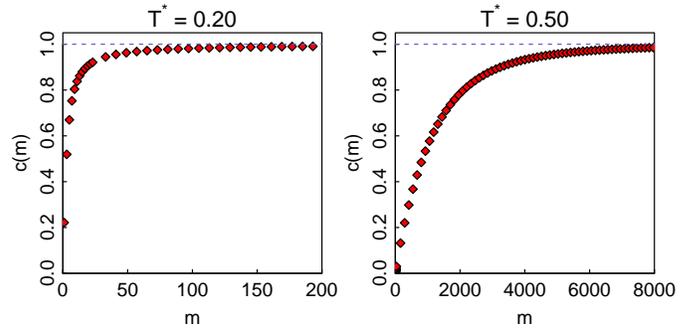}
\caption{Weight cumulant of eigen microstate in 1d-Ising model.}
\label{accumulaton1}
\end{figure}

In Fig. \ref{accumulaton1}, the weight cumulants of 1d-Ising model are plotted. At $T^*=0.2$, the cumulant $c(m)$ reaches nearly $1$ at $m \approx 200$. So the original microstates are constituted actually by about $200$ eigen microstates. At $T^*=0.5$,  $c(m)$ becomes nearly $1$ at $m \approx 8000$, which is still much less than $M=2\times 10^4$. 

\subsection{Two-dimensional Ising model}\label{sec:4}
In a two-dimensional (2d) Ising model with linear length $L$ and periodic boundary conditions, there are $N=L\times L$ spins in this system. With the nearest neighbor interaction $J$ and square lattice, this model has a ferromagnetic phase transition at the reduced temperature $T_c^*=k_B T_c/J=2/\ln (1+\sqrt{2})\approx 2.269$\cite{Onsager1944}.

\begin{figure}[H]
\centering
\includegraphics[scale=0.8]{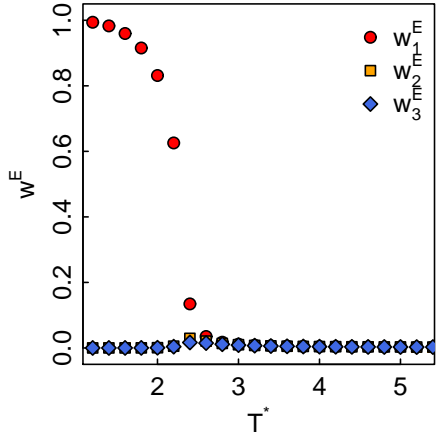}
\caption{The largest three weight factor of 2d-Ising model with $L=32$.  }
\label{fig:eigenvalues2}
\end{figure}

The microstates of the 2d-Ising model are simulated using the Wolff algorithm \cite{Wolff1989} also. In our simulations, we start with all spins aligned. The first $8000$ microstates are used to reach the equilibrium. From the subsequent microstates,   $M=2\times 10^4$ microstates at each temperature are taken at an interval of 250 MC steps. From the microstates, we can calculate the correlation matrix at first and then its eigenvalues and eigenvectors. The eigen microstates are obtained using the eigenvectors. 

The largest three eigenvalues around the critical point are presented in Fig. \ref{fig:eigenvalues2}. The normalized eigenvalue by $M$ is equivalent to the weight factor.
At temperatures above $T_c$, the largest three weight factors are quite small. There is no localization of eigen microstate. The weights of eigen microstates are distributed widely. At temperatures below $T_c$, the largest eigenvalues become finite. This indicates a condensation of the eigen microstate . There is now a phase transition, whose nature is characterized by the condensed eigen microstate.

\begin{figure}[H]
\centering
\includegraphics[scale=0.38]{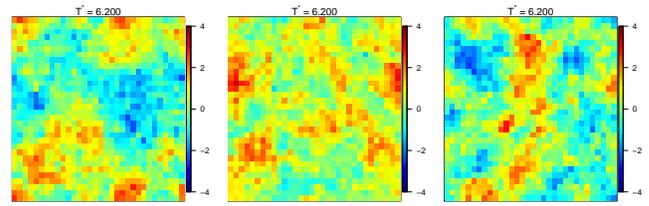}
\caption{The largest  eigen microstate $\sqrt{N}\bar{\pmb E}^1$(a), the second largest eigen microstate $\sqrt{N}\bar{\pmb E}^2$ (b), and the third largest eigen microstate $\sqrt{N}\bar{\pmb E}^3$ (c)  of 2d-Ising model above $T_c$.}
\label{microstates4}
\end{figure}

The largest three eigen microstates at $T^*=6.2$ are shown in Fig. \ref{microstates4}. The sizes of the spin clusters in the eigen microstates are much smaller than system size. 

\begin{figure}[H]
\centering
\includegraphics[scale=0.38]{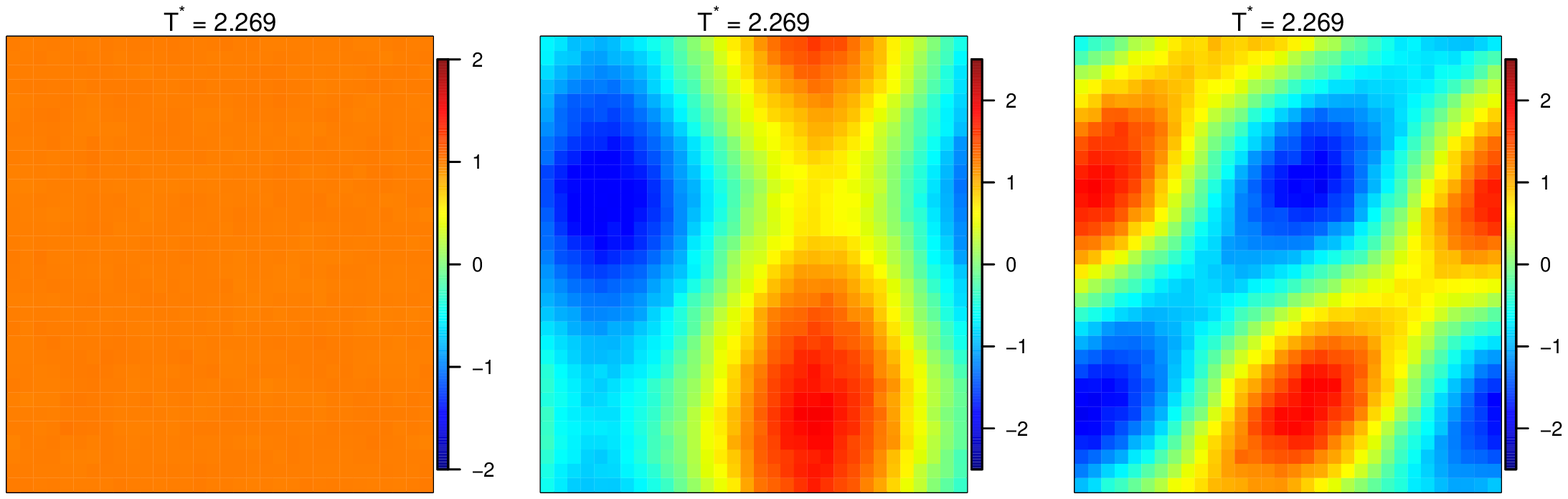}
\caption{The largest  eigen microstate $\sqrt{N}\bar{\pmb E}^1$(a), the second largest eigen microstate $\sqrt{N}\bar{\pmb E}^2$ (b), and the third largest eigen microstate $\sqrt{N}\bar{\pmb E}^3$ (c) of 2d-Ising model at $T_c^* \approx 2.269$.}
\label{microstates3}
\end{figure}

\begin{figure}[H]
\centering
\includegraphics[scale=0.38]{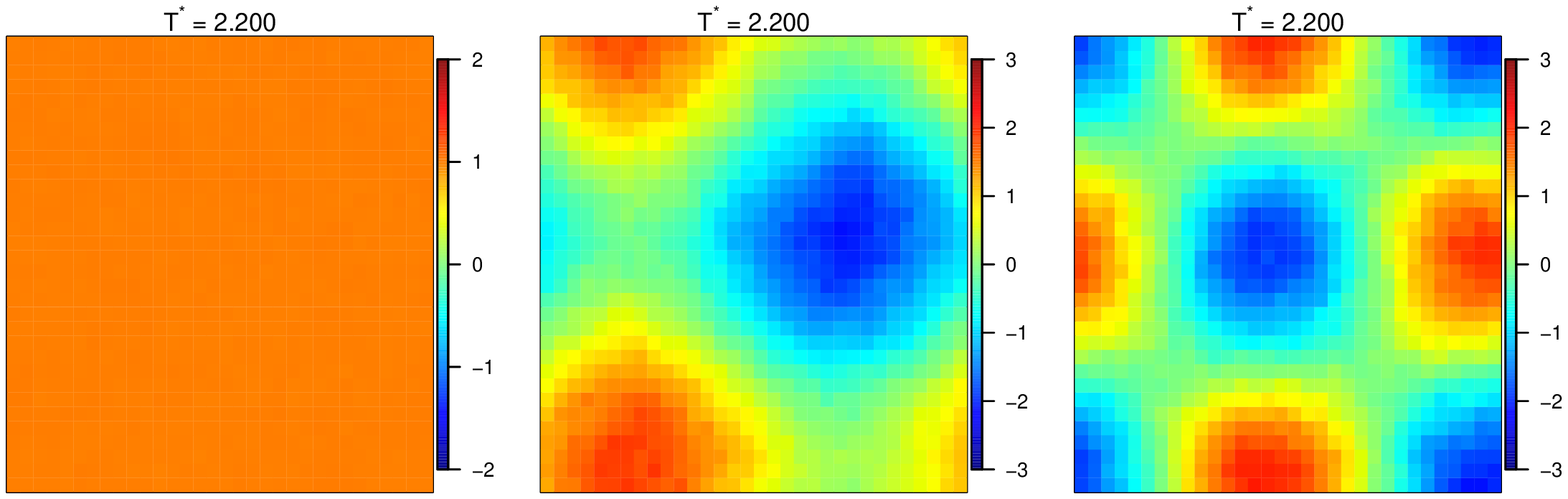}
\caption{The largest  eigen microstate $\sqrt{N}\bar{\pmb E}^1$(a), the second largest eigen microstate $\sqrt{N}\bar{\pmb E}^2$ (b), and the third largest eigen microstate $\sqrt{N}\bar{\pmb E}^3$ (c) of 2d-Ising model below $T_c$.}
\label{microstates2}
\end{figure}

\begin{figure}[H]
\centering
\includegraphics[scale=0.38]{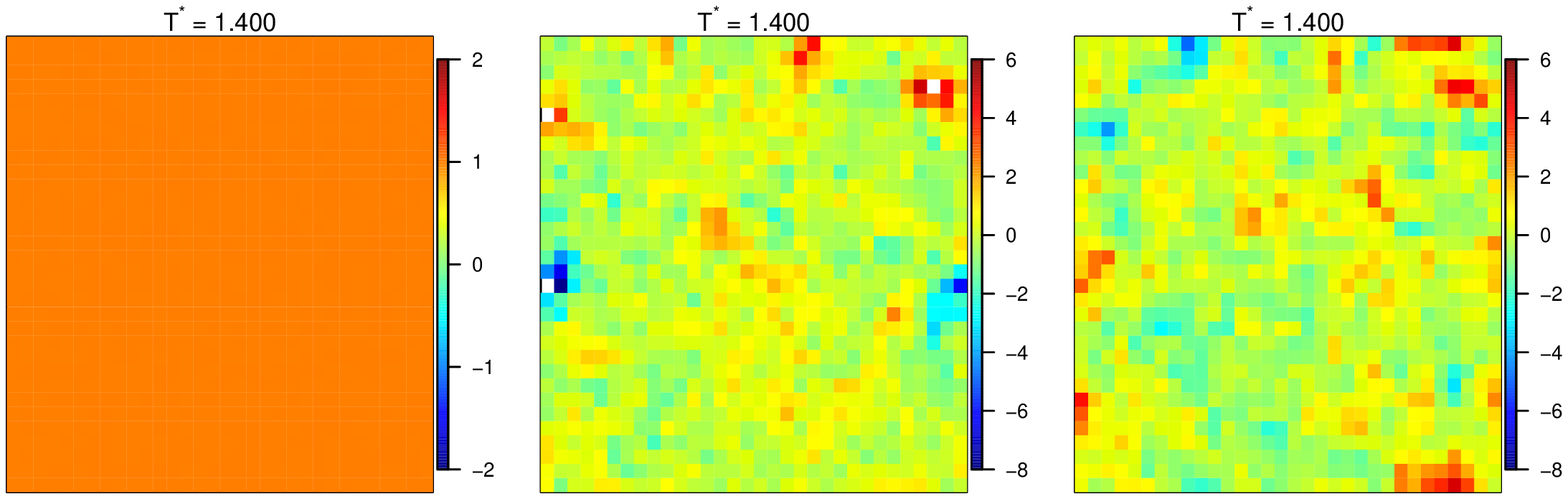}
\caption{The largest  eigen microstate $\sqrt{N}\bar{\pmb E}^1$(a), the second largest eigen microstate $\sqrt{N}\bar{\pmb E}^2$ (b), and the third largest eigen microstate $\sqrt{N}\bar{\pmb E}^3$ (c) of 2d-Ising model below $T_c$.}
\label{microstates1}
\end{figure}

In Fig. \ref{microstates3}, the eigen microstates of the largest three eigenvalues at $T^*_c \approx 2.269$ are presented. There is only one cluster in the eigen microstate of the largest eigenvalue. 

\begin{figure}[H]
\centering
\includegraphics[scale=0.55]{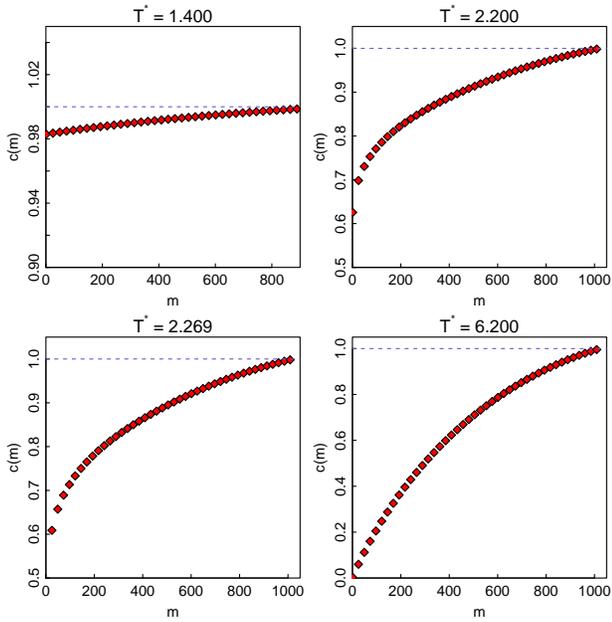}
\caption{Weight cumulant of eigen microstate in 2d-Ising model.}
\label{sum2}
\end{figure}

We plot in Fig. \ref{microstates2} the eigen microstates of the largest three eigenvalues at $T^*=2.2$. In the largest eigen microstate,  there is only one spin cluster. The large-$M$ limit of its weight factor is larger than $0.6$, as shown in Fig. \ref{sum2}. There is a ferromagnetic phase transition. The second largest eigen microstate has two clusters with opposite orientation. There are four spin clusters in the third largest eigen microstate.

With the further decrease of temperature, the largest eigen microstate ${\pmb E}_1$ with one spin cluster becomes more dominant. The weight factor of the largest eigen microstate at $T^*=1.4$ is larger than $0.98$, which can be seen in Fig. \ref{sum2}. Other eigen microstates have many small clusters, as shown in Fig. \ref{microstates1}.

The weight cumulants of eigen microstate at $T^*=1.4, 2.2, 2.269, 6.2$ are shown in Fig. \ref{sum2}. The weight cumulants reach nearly $1$ at $m \approx 1000$. The $2\times 10^4$ original microstates are composed of $1000$ eigen microstates approximately.

\section{Finite-size scaling of weight factor near critical point}\label{sec:5}
In the region near a critical point, thermodynamic functions of finite system are proposed to satisfy finite-size scaling relations \cite{Fisher1972Scaling,PhysRevB.30.322,Privman1991}. For the order parameter, its finite-size relation is
\begin{eqnarray}
\label{scalem}
    \left< m \right> (t,L) = L^{-\beta/\nu} f_m (tL^{1/\nu})\,
\end{eqnarray}
where $t=(T-T_c)/T_c$ is the reduced temperature, $\beta$ is the critical exponent of order parameter, and $\nu$ is the critical exponent of bulk correlation length $\xi = \xi_0 t^{-\nu}$. The scaling variable $tL^{1/\nu}$ is related to the size ratio $L/\xi$. 

The correlation length follows the finite-size scaling form $\xi (t,L) = L X(tL^{1/\nu})$ \cite{PhysRevB.30.322}. Recently, a finite-size scaling relation of correlation function was proposed \cite{Zhang2018}. Using this finite-size relation, the finite-size scaling form of correlation length can be naturally derived. For the principal fluctuation modes of complex system, there is also a finite-size scaling relation \cite{0253-6102-66-3-355}, which has been confirmed in 2d-Ising model. 

Basing on the relation between the weight factor of eigen microstate and the order parameter, we propose a finite-size scaling form

\begin{eqnarray}
\label{scalela}
    w_I^E (t,L) = L^{-2\beta/\nu} F_{w}^I (t L^{1/\nu})\;.
\end{eqnarray}

\begin{figure}[H]
\centering
\includegraphics[scale=0.7]{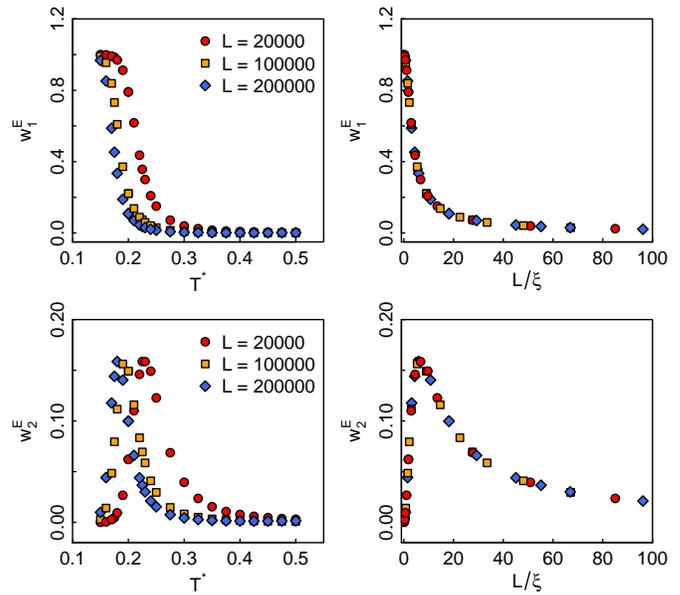}
\caption{The largest two weight factors of 1d-Ising model with respect to $T^*$ and $L/\xi$. }
\label{fig:1dscaling}
\end{figure}

In one-dimensional Ising model, the critical exponent $\beta = 0$ and $w_I^E (T,L) = F_w^{I} (L/\xi)$, where $\xi$ is given in Eq. \eqref{xi1}. We present the largest two weight factors with respect to $T^*$ and scaling variable $L/\xi$ in Fig. \ref{fig:1dscaling}. The finite-size scaling form of Eq. \eqref{scalela} is confirmed in 1d-Ising model.

\begin{figure}[H]
\centering
\includegraphics[scale=0.6]{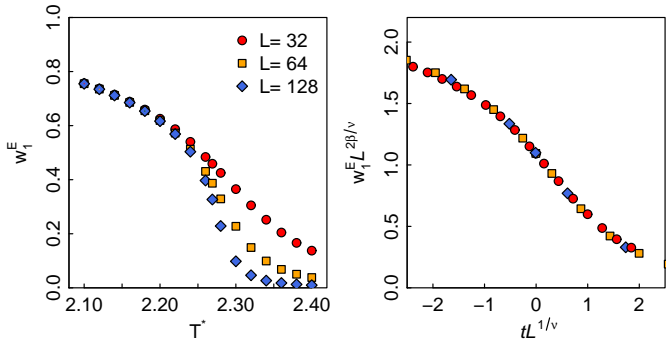}
\caption{The largest weight factor and its finite-size scaling form in 2d-Ising model.}
\label{eigen21}
\end{figure}

In two-dimensional Ising model, the critical exponents $\nu=1$ and $\beta=1/8$ \cite{Onsager1944}. On the left side of Fig. \ref{eigen21}, the largest weight factor $w_1^E$ of $L=32, 64, 128$ are plotted with respect to $T^*$. The finite-scaling form $w^E_1 L^{2\beta/\nu}$ is presented with respect to $tL^{1/\nu}$ on the right side. The different curves for different $L$ collapse together. The finite-size relation of Eq. \eqref{scalela} is confirmed in 2d-Ising model.

\begin{figure}[H]
\centering
\includegraphics[scale=0.6]{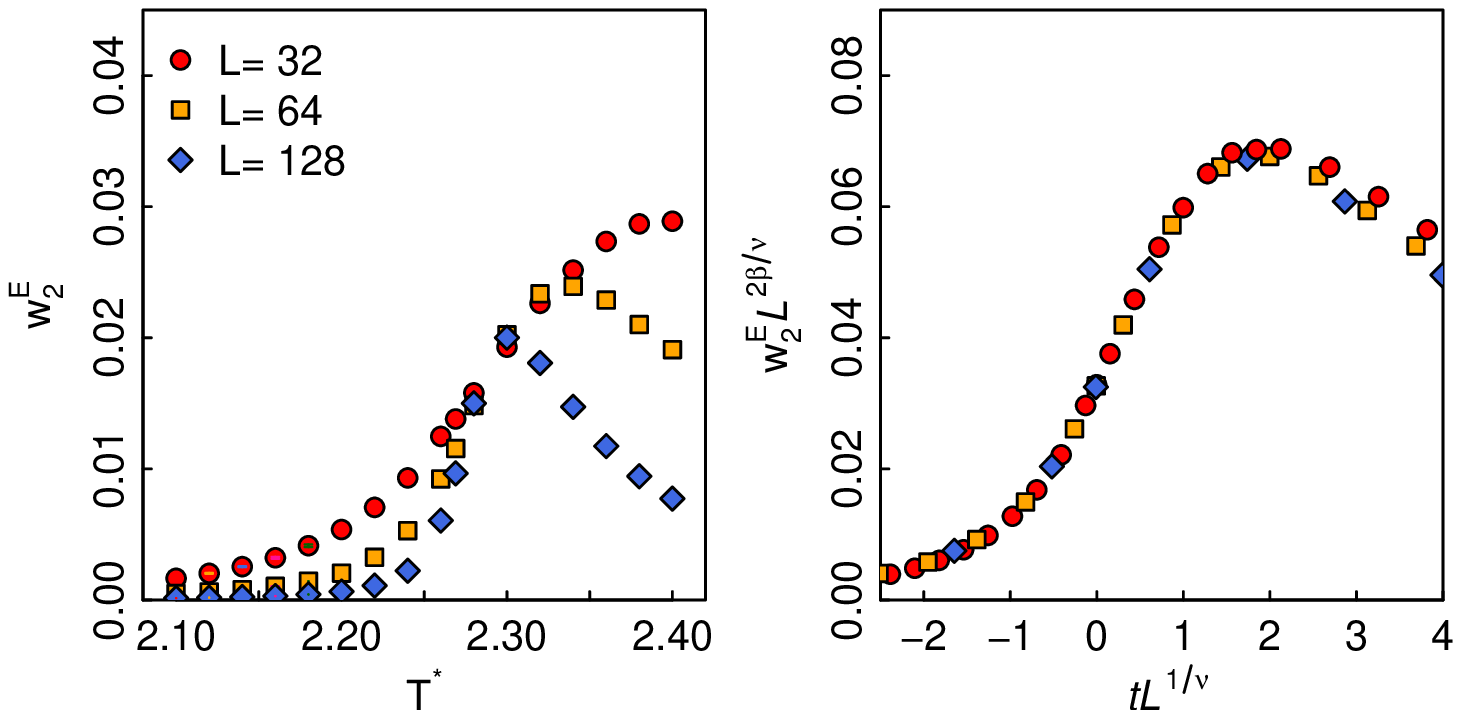}
\caption{The second largest weight factor and its finite-size scaling form in 2d-Ising model.}
\label{eigen22}
\end{figure}

\begin{figure}[H]
\centering
\includegraphics[scale=0.6]{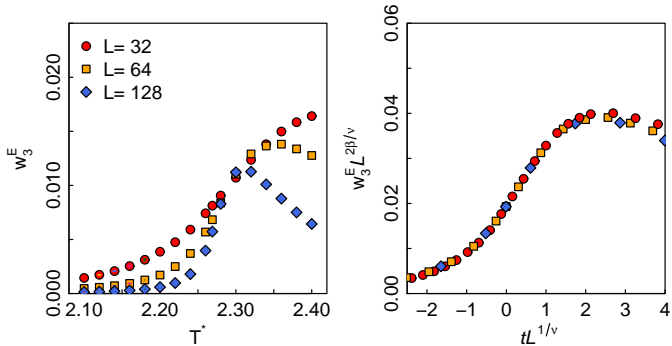}
\caption{The third largest weight factor and its finite-size scaling form in 2d-Ising model.}
\label{eigen23}
\end{figure}

In Figs. \ref{eigen22} and \ref{eigen23}, the second and third largest weight factors and their finite-size scaling form are presented.

After taking logarithm of Eq. \eqref{scalela}, we obtain
\begin{eqnarray}
\label{scaleln}
    \ln w_I^E (t,L) = -\left(2\beta/\nu\right) \ln L + \ln F_w^I (t L^{1/\nu})\;,
\end{eqnarray}
which depends on $\ln L$ linearly at $t=0$. This property can be used to determine the critical point and the critical exponent ratio $\beta/\nu$. 
\begin{figure}[H]
\centering
\includegraphics[scale=0.6]{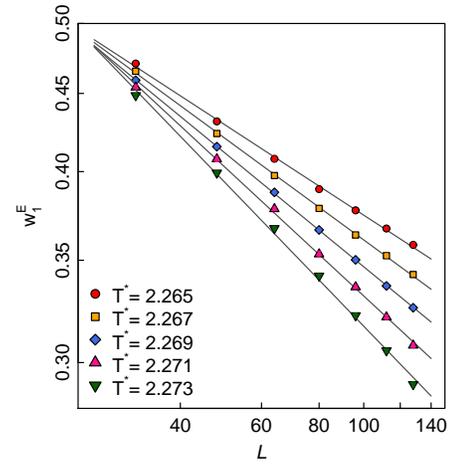}
\caption{Log-log plot of the largest normalized eigenvalue around $T_c$. }
\label{log-log}
\end{figure}
In Fig. \ref{log-log}, the log-log plot of the largest weight factor $w_1^E$ with respect to $L$ is presented at different reduced temperatures. The curves are curved upward at $T > T_c$ and downward at $T < T_c$. From the straight line at $T_c^* \approx 2.269$, we can obtain $2\beta/\nu=0.246(6)$, which is in agreement with the exact value $2\beta/\nu=1/4$ \cite{Onsager1944}.

We introduce the ratio of weight factor $R \equiv w_2^E/w_1^E$, which follows the finite-size scaling form
\begin{eqnarray}
    R (t,L)=F_w^2/F_w^1 = \tilde{R} (tL^{1/\nu})\;.
\end{eqnarray}
At the critical point, the ratio $R (0,L)= \tilde{R} (0)$ is independent of $L$. This can be used to determine the critical point also. 

\begin{figure}[H]
\centering
\includegraphics[scale=0.6]{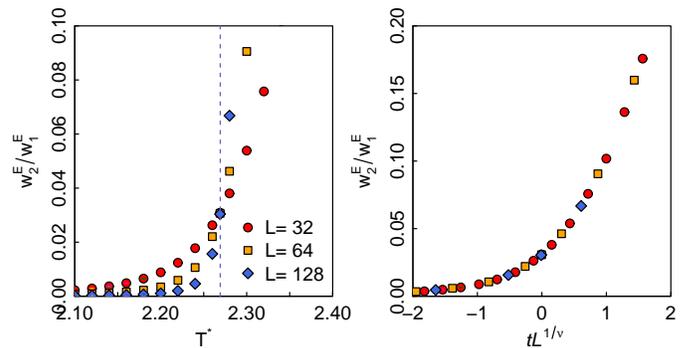}
\caption{Ratio of the largest eigenvalue to the second largest eigenvalue. }
\label{fixed}
\end{figure}
The ratio $R (t,L)$ of 2d-Ising model is plotted in Fig. \ref{fixed}. With $T^*$ as the variable, curves of different $L$ have a fixed point at $T_c$. Using $tL^{1/\nu}$ as the variable, the curves collapse together.

Therefore, the finite-size scaling relation of weight factor in Eq. \eqref{scalela} has been verified in one-dimensional and two-dimensional Ising models. We anticipate that this finite-size relation is valid for general complex systems.

\section{Conclusions}\label{sec:6}
In the phase space of a complex system, we introduce the eigen microstates of its statistical ensemble. The microstates of the complex system under some conditions can be obtained from computer simulations or experimental studies.  We introduce a correlation matrix with correlations between microstates as its elements. Using the eigenvectors of the correlation matrix, the eigen microstates of the ensemble can be defined. The normalized eigenvalues by the number of microstate $M$ can be considered as the weight factor in the ensemble of the corresponding eigen microstates.

In an ensemble without localization of microstate, weight factors of eigen microstate go to zero when $M \to \infty$. If the largest weight factor becomes finite in the limit $M \to \infty$, there is a condensation of the eigen microstate in the ensemble. This condensation indicates a phase transition with the new phase characterized by the eigen microstate corresponding to the finite weight factor. We propose a finite-size scaling relation of the weight factors near critical point using the critical exponents of order parameter and correlation length. 

The eigen microstates and their weight factors in an ensemble of one-dimensional and two-dimensional Ising models have been studied using Monte Carlo simulation. The condensation of eigen microstate in the one-dimensional Ising model appears when $T \to 0$. All spins of the condensed eigen microstate have the same orientation and there is a ferromagnetic phase transition. In two-dimensional Ising model, condensations of eigen microstate are found at the reduced temperatures $T^* < T_c^* =2/\ln (1+\sqrt{2})$. In the condensed eigen microstate, all spins have the same orientation and there is a ferromagnetic phase transition in two-dimensional Ising model. The finite-size scaling relation of weight factors is confirmed by the Monte Carlo simulation results of one-dimensional and two-dimensional Ising models. Further, we will study the eigen microstates and their weight factors of statistical ensemble for complex systems such as confined fluids \cite{Dong2018}, networks with long-range connections \cite{Yang2017}, and climate systems \cite{Fan2017}.

In the studies of phase transitions of complex systems, the definition of order parameter is sometimes a challenge. We can take collective motion as an example. Collective motion exists at almost every scale in nature, from unicellular organisms to bird flocks, fish schools, and human crowds \cite{Vicsek2012}. The simple model of collective motion was introduced by Vicsek and collaborators \cite{Vicsek}.The transition to collective motion in the Vicsek model (VM) is thought to be critical \cite{Vicsek} and discontinuous \cite{Gregoire2004}. More phases such as ordered "Toner-Tu" phase \cite{Toner1995} and band phases \cite{Bertin,Mishra,Ihle2011} are suggested to exist in the VM. A global understanding and the real order parameter of the transition to collective motion is still lacking. Using the method proposed here, we can determine the critical point, the order parameter, and the critical exponents of complex system at the same time. Our method of analysis is not restricted to systems in equilibrium.

\Acknowledgements{ This work was support by Key Research Program of Frontier Sciences, Chinese Academy of Sciences (Grant No. QYZD-SSW-SYS019). Our Monte Carlo simulations are supported by HPC Cluster of ITP-CAS. We are grateful to insight discussions with Prof. Jinghai Li.}

\InterestConflict{The authors declare that they have no conflict of interest.}



\end{multicols}
\end{document}